\documentclass[preprints,article,accept,oneauthor,pdftex]{Definitions/mdpi}

\providecommand{\gitinfo}{\today}

\firstpage{1}
\pubvolume{1}
\issuenum{1}
\articlenumber{0}
\pubyear{2021}
\copyrightyear{2021}
\datereceived{\gitinfo}
\dateaccepted{}
\datepublished{}
\hreflink{https://doi.org/}

\newcommand{\mat}[1]{{\boldsymbol{#1}}}

\newcommand{\bbR}{\mathbb{R}}
\newcommand{\cA}{\mathcal{A}}
\newcommand{\cB}{\mathcal{B}}
\newcommand{\cF}{\mathcal{F}}
\newcommand{\cM}{\mathcal{M}}
\newcommand{\cN}{\mathcal{N}}
\newcommand{\cS}{\mathcal{S}}
\newcommand{\exppr}{^{\mathrm{exp.}}}
\newcommand{\peakpr}{^{\mathrm{peak.}}}
\newcommand{\trsp}{^{\mathrm{T}}}
\newcommand{\udet}{_{\mathrm{det.}}}
\newcommand{\ufa}{_{\mathrm{f.a.}}}
\newcommand{\unifpr}{^{\mathrm{unif.}}}
\newcommand{\unosig}{_{\mathrm{no~signal}}}
\newcommand{\usig}{_{\mathrm{signal}}}
\newcommand{\uthr}{_{\star}}

\Title{Geometric Approach to Analytic Marginalisation of the Likelihood Ratio
  for Continuous Gravitational Wave Searches}

\TitleCitation{Geometric Approach to Analytic Marginalisation of the Likelihood
  Ratio for Continuous Gravitational Wave Searches}

\Author{Karl Wette $^{1}$\orcidA{}}

\AuthorNames{Karl Wette}

\AuthorCitation{Wette, K.}

\address{$^{1}$ \quad ARC Centre of Excellence for Gravitational Wave Discovery
  (OzGrav) and Centre for Gravitational Astrophysics, Australian National
  University, Canberra ACT 2601, Australia; karl.wette@anu.edu.au}

\corres{Correspondence: karl.wette@anu.edu.au}

\abstract{
  The likelihood ratio for a continuous gravitational wave signal is viewed
  geometrically as a function of the orientation of two vectors; one
  representing the optimal signal-to-noise ratio, the other representing the
  maximised likelihood ratio or $\cF$-statistic.  Analytic marginalisation over
  the angle between the vectors yields a marginalised likelihood ratio which is
  a function of the $\cF$-statistic.  Further analytic marginalisation over the
  optimal signal-to-noise ratio is explored using different choices of prior.
  Monte-Carlo simulations show that the marginalised likelihood ratios have
  identical detection power to the $\cF$-statistic.  This approach demonstrates
  a route to viewing the $\cF$-statistic in a Bayesian context, while retaining
  the advantages of its efficient computation.
}


\begin{document}

\section{Introduction}\label{sec:introduction}

Continuous gravitational waves are, at best, weak signals relative to the
sensitivity of current-generation interferometric
detectors~\cite{LIGO2015:AdvLIG, AcerEtAl2018:SttAdvVr,
  KAGRLIGOVirg2020:PrObLclGrvTrALAVK}.  Searches of data from the LIGO and Virgo
observatories, most recently from their
2nd~\cite{LIGOVirg2019:ASCntGrvWIsNtSUAdLOD, LIGOVirg2019:NrrSrGrvWKPlUSLObR,
  LIGOVirg2019:SGrvWScXSAdLObsRImHMM, LIGOVirg2019:SrGrvWvKPlTHrm20152017LD,
  PaloEtAl2019:DCnsUlBMSrCnGrvW, CovaSint2020:FASCntGrvSgUnNSBSUAdLD,
  DergPapa2020:RsFASCnGrvWSmlSr, FesiPapa2020:FrSrRmGrvWPJ05,
  LindOwen2020:DrSCntGrvWTSpRDALSObR, MiddEtAl2020:SGrvWFLMXBnSAdLObsRImHMM,
  MillEtAl2020:SGrvW12YSprRmHMMAdLSObR, PiccEtAl2020:DrSCntGrvSGlCALSObR,
  SunEtAl2020:SrUltBsCyXAdvL, ZhanEtAl2021:SrCntGrvWvScXLOD,
  JoneSun2021:SCntGrvWFmBSAdLObRHMM, BeniEtAl2021:SrCntGrvWTHSrUHdMM, 1HzSearch}
and 3rd observing runs~\cite{LIGOVirg2020:GrvCnsEqElMlP,
  LIGOVirg2021:ASEOLDCntGrvSgUnNtSBS}, have yet to make a first detection.
Theoretical modelling of rapidly-rotating, non-axisymmetric neutron stars -- the
most likely source of continuous waves -- predict a wide range of possible
signal strengths~\cite{BonaGour1996:GrvWPlEmMgFIDst,
  UshoEtAl2000:DfrAcNtSCGrvWEm, Owen2005:MxElDfrCmSEEqtS,
  HaskEtAl2008:MdlMgnDfNtSt, GlamEtAl2012:GrvWvClrMnNtS,
  JohnOwen2013:MxElDfrRltSt, WoanEtAl2018:EvMnEllMllPl,
  OsboJone2020:GrvWMgnInThNSMn}.  Optimally-sensitive data analysis techniques
are therefore important.

Given an assumed signal model -- a quasisinusoid which evolves with the rotation
frequency of the neutron star, and is modulated by the relative motion between
the star and an Earth-based detector -- a matched filter can be constructed to
achieve maximum \emph{detection power}, in the
Neyman--Pearson~\cite{NeymPear1933:PrMEfTsSttHyp} sense of maximising the
probability of detection (true positive) at a given probability of false alarm
(false positive).  Furthermore, as first shown
in~\cite{JaraEtAl1998:DAnGrvSgSpNSSDtc}, the matched filter likelihood ratio can
be analytically maximised over four \emph{amplitude parameters}
$\cA_1, \cA_2, \cA_3, \cA_4$, resulting in the well-known $\cF$-statistic.

The Bayesian approach to signal detection and parameter inference has become
central to gravitational-wave
astronomy~\cite[e.g.][]{ThraTalb2019:IntBInGrvAPEsMSHrM}.  It was recognised
in~\cite{SearEtAl2008:RbByDtcUnmBr, Sear2008:MntByTcGrvWBDAn} that maximisation
over signal parameters can bias detection statistics: from the Bayesian
viewpoint, maximisation implicitly assumes prior probabilities for the maximised
parameters, which may not be physically motivated.

In~\cite{PrixKris2009:TrSCnGrvWBVMxmSt} the $\cF$-statistic is shown to possess
such a bias due to analytic maximisation over the four amplitude parameters.
The $\cA_1, \cA_2, \cA_3, \cA_4$ are functions of four physical parameters of
the continuous wave signal model: the overall signal strength $h_0$; the
inclination $\iota$ and polarisation $\psi$ angles, which orient the neutron
star rotation axis relative to the observer; and the signal phase $\phi_0$ at
some reference time.  Given no prior knowledge of the orientation of the neutron
star, or the signal phase, one would assume uniform priors on $\cos\iota$,
$\psi$, and $\phi_0$; and the absence of detections of continuous wave to date
is consistent with a choice of prior on $h_0$ which prefers weaker signals to
stronger ones.  The $\cF$-statistic, however, implicitly adopts priors which
prefer stronger signals (i.e.\ larger $h_0$) compared to weaker ones.  It is
also biased in favour of \emph{linearly polarised} signals where
$\cos\iota \sim 0$ (i.e.\ the neutron star is viewed ``edge-on'' with the
rotation axis at right angles to the line of sight) compared to \emph{circularly
  polarised} signals where $|\cos\iota| = 1$ (i.e.\ the neutron star is viewed
``face-on'' with the rotation axis parallel to the line of sight).

By instead marginalising the likelihood ratio over $h_0$, $\cos\iota$, $\psi$,
and $\phi_0$ with physically-motivated
priors,~\cite{PrixKris2009:TrSCnGrvWBVMxmSt} introduced the $\cB$-statistic, a
Bayesian alternative to the $\cF$-statistic.  Monte-Carlo simulations were
performed to estimate the \emph{receiver-operator curve}, which plots the
probability of detection against the probability of false alarm.  The
$\cB$-statistic was found to be a more powerful detection statistic than the
$\cF$-statistic, assuming a signal population where the distributions of
$\cos\iota$, $\psi$, and $\phi_0$ are consistent with the $\cB$-statistic
priors~\cite{Sear2008:MntByTcGrvWBDAn}.

A practical downside of the $\cB$-statistic is that, to date, a convenient
analytic expression for the marginalised likelihood ratio has not been found,
and therefore the marginalisation must be performed by numerical integration.
This puts the $\cB$-statistic at a disadvantage with respect to the
$\cF$-statistic, for which computationally efficient implementations
exist~\cite{JaraEtAl1998:DAnGrvSgSpNSSDtc, Prix2010:FsImpCmp,
  PateEtAl2010:ImpBrRsCnWSGrvWD, PoghEtAl2015:ArImpPrlSfSPGrWS}.  Past work has
sought to address this issue though transformation of the amplitude parameters
to new coordinate systems, and approximations to the marginalisation integrals
in various limits~\cite{Derg2012:LsChrSrSWllSg, WhelEtAl2014:NCrdAmPrSCnGrvW,
  DhurEtAl2017:MrgLkFnMGrvWSr, BeroWhel2019:AAppBDtStCnGrvW}.

This paper presents an alternative route to marginalising the likelihood ratio
for continuous gravitational wave searches.  A geometric view of the likelihood
ratio is presented in Sec.~\ref{sec:geom-view-likel}, which permits analytic
marginalisation over its parameters in Sec.~\ref{sec:analyt-marg-likel}.
Receiver-operator curves for the marginalised likelihood ratio are presented in
Sec.~\ref{sec:rece-oper-curv}, and a discussion in Sec.~\ref{sec:discussion}
concludes the paper.

\section{Geometric View of the Likelihood Ratio}\label{sec:geom-view-likel}

Gravitational waves detectors measure strain, the differential displacement
between test particles due to a passing gravitational wave.  The strain due to a
continuous wave signal may be written as~\cite{JaraEtAl1998:DAnGrvSgSpNSSDtc}
\par
\begin{equation}
  \label{eq:hoft}
  h(t, \cA, p) = \vec{\cA} \cdot \vec{h}(t, p) \,,
\end{equation}
\par\noindent
where $\vec{\cA} \in \bbR^4$ is a vector of the amplitude parameters, and
$\vec{h}(t, p) \in \bbR^4$ is a vector of time-dependent basis
functions.\footnote{The dot product $\alpha \cdot \beta$ henceforth denotes the
  contraction of the last index of the tensor $\alpha$ with the first index of
  the tensor $\beta$.}  Additional parameters $p$ of $\vec{h}$ encode the phase
modulation of the continuous wave signal: these typically include Taylor
coefficients of the evolution of the gravitational wave frequency, the position
of the neutron star in the sky, and if necessary parameters of the orbit of the
neutron star around a companion.

The likelihood ratio for continuous waves arises from considering two
hypotheses: that the data $x(t)$ consists only of Gaussian stationary noise, with
single-sided power spectral density $\cS$; or that the data additionally
contains a signal specified by Eq.~\eqref{eq:hoft}.  The log-likelihood ratio
between the two hypotheses is then~\cite{JaraEtAl1998:DAnGrvSgSpNSSDtc, Prix2007:SrCnGrvWMMltFs}
\par
\begin{equation}
  \label{eq:logLambda-full}
  \ln \Lambda(x; \cA, p) = \vec{\cA} \cdot \vec{X}(x; p) - \frac{1}{2} \vec{\cA} \cdot \mat{\cM} \cdot \vec{\cA} \,.
\end{equation}
\par\noindent
A search for continuous wave is performed by repeated computation of
Eq.~\eqref{eq:logLambda-full} for different choices of $p$, corresponding to
different choices of signal hypothesis.  Typically, a fixed set of $p$ called a
\emph{template bank} is constructed, in such a way as ensure any signal in
$x(t)$ matches at least one of the signal hypotheses with low loss in signal-to-noise
ratio, typically $\lesssim 30$\%~\cite{BradEtAl1998:SrcPrdSrLI}.
A metric on the parameter space of $p$ is often used in constructing
template banks~\cite{BalaEtAl1996:GrvWClsBDStMCEsPr, Owen1996:STmGrvWInsBnCTmS,
  Prix2007:SrCnGrvWMMltFs, WettPrix2013:FPrmMtASrGrvPl}.

The elements of the vector $\vec{X}(x; p) \in \bbR^4$ in
Eq.~\eqref{eq:logLambda-full} are inner products (normalised by $\cS$) of the
data $x(t)$ with the basis functions $\vec{h}(t, p)$. The elements of the matrix
$\mat{\cM} \in \bbR^4 \otimes \bbR^4$ are inner products of the $\vec{h}(t, p)$ with
each other.  The typical time-span of data searched for continuous waves (days
to years) far exceeds the time-scale of oscillations in $\vec{h}(t, p)$ due to
the gravitational wave frequency ($\sim 1$--$10^3$~Hz); as a result, some inner
products between the $\vec{h}(t, p)$ quickly average to zero.  The remaining
non-zero elements of $\mat{\cM}$ are~\cite{JaraEtAl1998:DAnGrvSgSpNSSDtc,
  KrolEtAl2004:OptFltLIDt, WhelEtAl2008:ImSGlWhtBMLDCh1UFstTB}
\par
\begin{equation}
  \label{eq:M-matrix}
  \mat{\cM} = \frac{1}{2}
  \begin{pmatrix}
    A & C & 0 & E \\
    C & B & -E & 0 \\
    0 & -E & A & C \\
    E & 0 & C & B
  \end{pmatrix} \,.
\end{equation}
\par\noindent
The element $E = 0$ under the assumption that the gravitational wavelength is
much larger than the size of the detector; this holds for terrestrial
gravitational-wave interferometers, though not for proposed space-based
detectors~\cite{KrolEtAl2004:OptFltLIDt, WhelEtAl2008:ImSGlWhtBMLDCh1UFstTB}.
The elements $A$, $B$, and $C$ can be expressed as inner products between two
functions $a(t, p)$ and $b(t, p)$, which are related to the response of the
gravitational wave detector to the two fundamental polarisations -- ``plus'' and
``cross'' -- of gravitational waves in general relativity.

The matrix $\mat{\cM}$ is symmetric and positive
definite~\cite{JaraEtAl1998:DAnGrvSgSpNSSDtc}.  It follows that its four leading
principal minors $D_1$, $D_2$, $D_3$, and $D_4$ are all strictly positive:
\par
\begin{subequations}
  \label{eq:M-lead-prin-minors}
  \begin{align}
    D_1 &= \det \frac{1}{2} \begin{pmatrix}
      A
    \end{pmatrix} = \frac{1}{2} A > 0 \,, \\
    D_2 &= \det \frac{1}{2} \begin{pmatrix}
      A & C \\
      C & B
    \end{pmatrix} = \frac{1}{4} ( A B - C^2 ) > 0 \,, \\
    D_3 &= \det \frac{1}{2} \begin{pmatrix}
      A & C & 0 \\
      C & B & -E \\
      0 & -E & A
    \end{pmatrix} = \frac{1}{8} A ( A B - C^2 - E^2 ) > 0 \,, \\
    D_4 &= \det \mat{\cM} = \frac{1}{16} ( A B - C^2 - E^2 )^2 > 0 \,.
  \end{align}
\end{subequations}
\par\noindent
It also follows that $\mat{\cM}$ possesses a Cholesky decomposition: a lower
triangular matrix $\mat{\cN} \in \bbR^4 \otimes \bbR^4$ such that
$\mat{\cM} = \mat{\cN} \mat{\cN}\trsp$, where $\mat{\cN}\trsp$ is the transpose of $\mat{\cN}$. The elements
of $\mat{\cN}$ are given in terms of the elements of $\mat{\cM}$ and the leading principal
minors $D_2$ and $D_3$:
\par
\begin{equation}
  \label{eq:N-matrix}
  \mat{\cN} = \begin{pmatrix}
    \sqrt{ A / 2 } & 0 & 0 & 0 \\
    C / \sqrt{ 2 A } & \sqrt{ 2 D_2 / A } & 0 & 0 \\
    0 & - E / \sqrt{ 8 D_2 / A } & \sqrt{ D_3 / D_2 } & 0 \\
    E / \sqrt{ 2 A } & - C E / \sqrt{ 8 A D_2 } & C \sqrt{ D_3 / (A^2 D_2) } & 2 \sqrt{ D_3 / A^2 }
  \end{pmatrix} \,.
\end{equation}

Define the vectors
\par
\begin{align}
  \label{eq:b-vector}
  \vec{b} &= \mat{\cN}\trsp \cdot \vec{\cA} = \begin{pmatrix}
    ( \cA_1 A + \cA_2 C + \cA_4 E ) / \sqrt{ 2 A } \\
    ( 4 \cA_2 D_2 - \cA_3 A E - \cA_4 C E ) / \sqrt{ 8 A D_2 } \\
    ( \cA_3 + \cA_4 ( C / A ) ) \sqrt{ D_3 / D_2 } \\
    ( 2 \cA_4 \sqrt{ D_3 } ) / A
  \end{pmatrix} \,, \\
  \label{eq:y-vector}
  \vec{y}(x; p) &= \mat{\cN}^{-1} \cdot \vec{X}(x; p) = \begin{pmatrix}
    X_1 \sqrt{ 2 / A } \\
    ( A X_2 - C X_1 ) / \sqrt{ 2 A D_2 } \\
    ( ( 4 D_2 X_3 + A E X_2 - C E X_1 ) / ( 4 D_3) ) \sqrt{ D_3 / D_2 } \\
    ( A X_4 - C X_3 - E X_1 ) / (2 \sqrt{ D_3 } )
  \end{pmatrix} \,.
\end{align}
\par\noindent
The log-likelihood ratio of Eq.~\eqref{eq:logLambda-full} can then be re-expressed as
\par
\begin{equation}
  \label{eq:logLambda-b-y}
  \ln \Lambda(x; \cA, p) = \vec{b} \cdot \vec{y}(x; p) - \frac{1}{2} \| \vec{b} \|^2 \,,
\end{equation}
\par\noindent
where $\| \vec{b} \|^2 \equiv \vec{b} \cdot \vec{b}$ defines the vector norm.
The lengths of the vectors $\vec{b}$ and $\vec{y}(x; p)$ are related to two well-known
quantities.  The length of $\vec{b}$ is proportional to the optimal signal-to-noise
ratio of the matched filter~\cite[cf.][Eq.~(24)]{Prix2007:SrCnGrvWMMltFs}:
\par
\begin{equation}
  \label{eq:b-length}
  2 \| \vec{b} \|^2 \equiv \rho^2 = A (\cA_1^2 + \cA_3^2) + B (\cA_2^2 + \cA_4^2) + 2 C ( \cA_1 \cA_2 + \cA_3 \cA_4) + 2 E ( \cA_1 \cA_4 - \cA_2 \cA_3 ) \,.
\end{equation}
\par\noindent
The length of $\vec{y}(x; p)$ is proportional to the $\cF$-statistic\footnote{It
  is common in the literature to quote values of twice the $\cF$-statistic,
  i.e.\ $2\cF$. This convention is not followed in this
  paper, however.}~\cite[cf.][Eq.~(19)]{Prix2007:SrCnGrvWMMltFs}:
\par
\begin{equation}
  \label{eq:y-length}
  \frac{1}{2} \| \vec{y} \|^2 \equiv \cF = \frac{ A (X_2^2 + X_4^2) + B (X_1^2 + X_3^2) - 2 C ( X_1 X_2  + X_3 X_4 ) - 2 E ( X_1 X_4 - X_2 X_3 ) }{ A B - C^2 - E^2 } \,.
\end{equation}
\par\noindent
Let
\par
\begin{equation}
  \label{eq:kappa}
  \kappa = \frac{ \vec{b} \cdot \vec{y}(x; p) } { \| \vec{b} \| \| \vec{y} \| }
\end{equation}
\par\noindent
be the cosine of the angle between $\vec{b}$ and $\vec{y}(x; p)$.  Substitution
of Eqs.~\eqref{eq:b-length},~\eqref{eq:y-length}, and~\eqref{eq:kappa} into
Eq.~\eqref{eq:logLambda-b-y} gives
\par
\begin{equation}
  \label{eq:logLambda-rho-kappa-F}
  \ln \Lambda(x; \rho, \kappa, \cF) = \rho \kappa \sqrt{ \cF } - \frac{1}{4} \rho^2 \,.
\end{equation}

\begin{figure}
\centering
\includegraphics[width=10.5 cm]{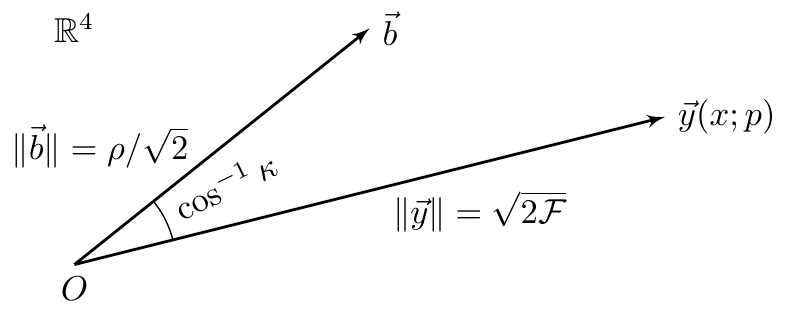}
\caption{\label{fig:b-y-vectors}
  Schematic of the vectors $\vec{b}$ [Eq.~\eqref{eq:b-vector}] and $\vec{y}(x; p)$ [Eq.~\eqref{eq:y-vector}], their lengths [Eqs.~\eqref{eq:b-length} and~\eqref{eq:y-length}], and the cosine
  $\kappa$ of the angle between them [Eq.~\eqref{eq:kappa}].
}
\end{figure}

As shown in Fig.~\ref{fig:b-y-vectors}, the log-likelihood ratio may be viewed
geometrically as a function of the relative orientation of two vectors.  One
vector, $\vec{y}(x; p)$, is a function of the data $x(t)$, and represents the
matched filter; the other vector, $\vec{b}$, represents the expected
signal-to-noise ratio.  Maximisation of the log-likelihood ratio with respect to
$\vec{\cA}$ is equivalent to aligning $\vec{b}$ and $\vec{y}(x; p)$:
maximising Eq.~\eqref{eq:logLambda-rho-kappa-F} with respect to $\rho$ gives
\par
\begin{equation}
  \label{eq:logLambda-rho-max-kappa-F}
  \max_{\rho} \ln \Lambda(x; \rho, \kappa, \cF) = \cF \kappa^2 \quad \text{at} \quad \rho = 2 \kappa \sqrt{ \cF } \,,
\end{equation}
\par\noindent
and Eqs.~\eqref{eq:logLambda-rho-kappa-F} and.~\eqref{eq:logLambda-rho-max-kappa-F} are
maximised when $\kappa = 1$, i.e.\ when $\vec{b}$ and $\vec{y}(x; p)$ are
parallel:
\par
\begin{equation}
  \label{eq:logLambda-rho-max-kappa-max-F}
  \max_{\rho, \kappa} \ln \Lambda(x; \rho, \kappa, \cF) = \cF \,.
\end{equation}

\section{Analytic Marginalisation of the Likelihood Ratio}\label{sec:analyt-marg-likel}

Instead of maximising the likelihood ratio with respect to $\kappa$ and $\rho$,
one could marginalise over these parameters with suitable priors.
Marginalisation over $\kappa$ is performed in Sec.~\ref{sec:marg-over-kappa},
followed by marginalisation over $\rho$, considering different choices of prior,
in Sec.~\ref{sec:marg-over-rho}.

\subsection{Marginalisation over $\kappa$}\label{sec:marg-over-kappa}

In the absence of a deeper understanding of the relationship between $\vec{b}$
and $\vec{y}(x; p)$, it is not unreasonable to adopt a prior on $\kappa$ that
assumes no preferred orientation between the two vectors.  The prior on $\kappa$
is then given by the distribution of $\vec{u} \cdot \vec{v}$, where
$\vec{u} \in \bbR^4$ and $\vec{v}\in \bbR^4$ are unit vectors uniformly
distributed on the 3-sphere $S^3 \subset \bbR^4$.

By invoking spherical symmetry, one can without loss of generality fix one
vector, say $\vec{u} = (1,0,0,0)$.  The problem then reduces to finding the
distribution of $\vec{u} \cdot \vec{v} = v_1$.  It is well
known~\cite{Mars1972:ChsPnSrfSp} that a vector uniformly distributed on the
$(d-1)$-sphere $S^{d-1} \subset \bbR^d$ may be found by generating a vector
$\vec{z} \in \bbR^d$ whose elements are independent standard normal variates,
then normalising $\vec{z}$ to unit length. Applying this procedure to $\vec{v}$,
the square of its first element is therefore
\par
\begin{equation}
  \label{eq:dist-v_1-sqr}
  v_1^2 = \frac{ z_1^2 }{ z_1^2 + z_2^2 + z_3^2 + z_4^2 } \,.
\end{equation}
\par\noindent
The distributions of $z_1^2$ and $z_2^2 + z_3^2 + z_4^2$ are chi-squared
distributions with 1 and 3 degrees of freedom respectively.  It follows that the
distribution of $\kappa^2 \sim v_1^2$ is a beta distribution with parameters
$\alpha = 1/2$, $\beta = 3/2$:
\par
\begin{equation}
  \label{eq:prior-kappa-sqr}
  p(\kappa^2) = \frac{ 2 \sqrt{ 1 - \kappa^2 } }{ \pi | \kappa | } \,, \quad 0 \le \kappa^2 \le 1 \,.
\end{equation}
\par\noindent
To find the distribution of $\kappa$, perform a change of variables and expand
the range of the distribution to $[-1,1]$:
\par
\begin{align}
  p(\kappa) = p(\kappa^2) \left| \frac{ d (\kappa^2) }{ d \kappa } \right|
  &= \frac{ 4 \sqrt{ 1 - \kappa^2 } }{ \pi } \,, \quad -1 \le \kappa \le 0 \text{~or~} 0 \le \kappa \le 1 \,; \\
  \label{eq:prior-kappa}
  &= \frac{ 2 \sqrt{ 1 - \kappa^2 } }{ \pi } \,, \quad -1 \le \kappa \le 1 \,.
\end{align}

\begin{figure}
\centering
\includegraphics[width=10.5 cm]{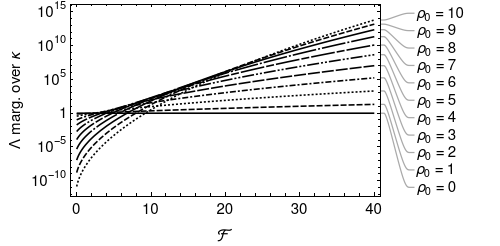} \\   
\includegraphics[width=10.5 cm]{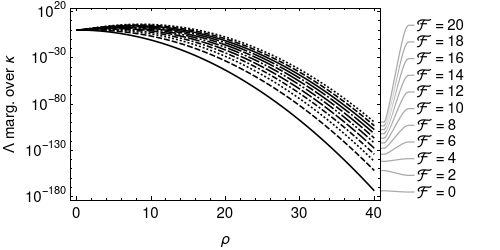}   
\caption{\label{fig:Lambda-marg-kappa}
  The likelihood ratio marginalised over $\kappa$ [Eq.~\eqref{eq:Lambda-rho-F}].
  Top: as a function of $\cF$ for fixed values of $\rho$.
  Bottom: as a function of $\rho$ for fixed values of $\cF$.
}
\end{figure}

Marginalisation of the likelihood ratio, in the form of
Eq.~\eqref{eq:logLambda-rho-kappa-F}, over $\kappa$ with the prior of
Eq.~\eqref{eq:prior-kappa} gives the analytic expression
\par
\begin{equation}
  \label{eq:Lambda-rho-F}
  \begin{split}
    \Lambda(x; \rho, \cF) &= \int_{-1}^{1} d\kappa \, p(\kappa) \Lambda(x; \rho, \kappa, \cF) \\
    &= \frac{ 2 }{ \rho \sqrt{\cF} } I_1 ( \rho \sqrt{\cF} ) e^{ -\rho^2 / 4 } \,,
  \end{split}
\end{equation}
\par\noindent
where $I_n$ is the modified Bessel function of the first kind of order $n$.
This function of $\rho$ and $\cF$ is plotted in
Fig.~\ref{fig:Lambda-marg-kappa}.  When $\rho$ is fixed, $\Lambda(x; \rho, \cF)$
is a monotonically increasing function of $\cF$.  When $\cF$ is fixed,
$\Lambda(x; \rho, \cF)$ monotonically decreases as a function of $\rho$ for
$\cF \le 2$, but achieves a local maximum at some $\rho > 0$ for $\cF > 2$.

\subsection{Marginalisation over $\rho$}\label{sec:marg-over-rho}

The marginalised likelihood ratio of Eq.~\eqref{eq:Lambda-rho-F} may be
further analytically marginalised over $\rho$, depending on its choice of
prior. For example, the choice of a uniform (improper) prior on $\rho$,
\par
\begin{equation}
  \label{eq:prior-rho-unif}
  p\unifpr(\rho) = 1 \,,
\end{equation}
\par\noindent
leads to
\par
\begin{equation}
  \label{eq:Lambda-F-unif}
  \begin{split}
    \Lambda\unifpr(x; \cF) &= \int_{0}^{\infty} d\rho \, p\unifpr(\rho) \Lambda(x; \rho, \cF) \\
    &= \sqrt{ \pi } \left[ I_0 \left( \frac{ \cF }{ 2 } \right) - I_1 \left( \frac{ \cF }{ 2 } \right) \right] e^{ \cF / 2 } \,.
  \end{split}
\end{equation}
\par
\begin{figure}[p]
\centering
\includegraphics[width=10.5 cm]{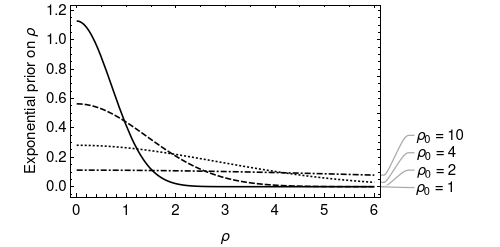} \\   
\includegraphics[width=10.5 cm]{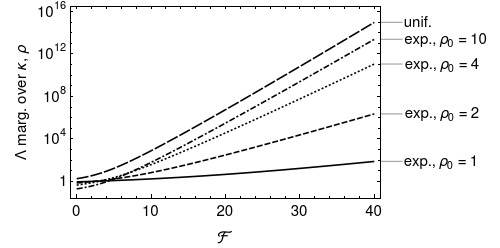}   
\caption{\label{fig:exp-prior-Lambda}
  Top: the exponential prior on $\rho$ [Eq.~\eqref{eq:prior-rho-exp}] as a
  function of $\rho$, for choices of the parameter $\rho_0$.  Bottom: the
  likelihood ratio marginalised over $\kappa$ and $\rho$ as a function of $\cF$,
  with the exponential prior on $\rho$ [Eq.~\eqref{eq:Lambda-F-exp}] for choices
  of $\rho_0$, and with the uniform prior [Eq.~\eqref{eq:Lambda-F-unif}].
}
\end{figure}

\begin{figure}[p]
\centering
\includegraphics[width=10.5 cm]{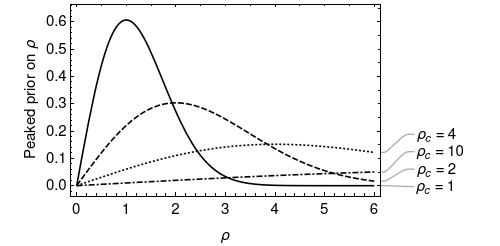} \\   
\includegraphics[width=10.5 cm]{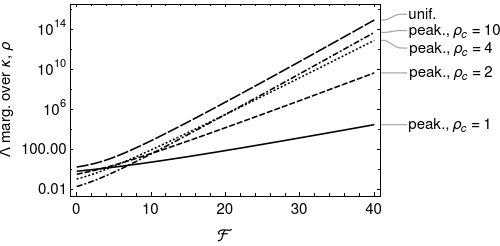}   
\caption{\label{fig:peak-prior-Lambda}
  Top: the peaked prior on $\rho$ [Eq.~\eqref{eq:prior-rho-peak}] as a
  function of $\rho$, for choices of the parameter $\rho_c$.  Bottom: the
  likelihood ratio marginalised over $\kappa$ and $\rho$ as a function of $\cF$,
  with the peaked prior on $\rho$ [Eq.~\eqref{eq:Lambda-F-peak}] for choices
  of $\rho_c$, and with the uniform prior [Eq.~\eqref{eq:Lambda-F-unif}].
}
\end{figure}

\noindent
This is a strictly increasing function of $\cF$, and is plotted in
Figs.~\ref{fig:exp-prior-Lambda} and~\ref{fig:peak-prior-Lambda}.

Another possible choice is an exponential prior on $\rho$:
\par
\begin{equation}
  \label{eq:prior-rho-exp}
  p\exppr(\rho) = \frac{ 2 }{ \rho_0 \sqrt{ \pi } } e^{ -( \rho / \rho_0 )^2 } \,,
\end{equation}
\par\noindent
with parameter $\rho_0$.  This choice of prior is consistent with the assumption
that the signal-to-noise ratio of continuous wave signals is weak, with the most
likely value at $\rho = 0$, and most values at $\rho \lesssim \rho_0$.
Fig.~\ref{fig:exp-prior-Lambda} plots the exponential priors for choices of the
parameter $\rho_0$; larger values of $\rho_0$ lower the peak at $\rho = 0$ and
flatten out the distribution.  Marginalisation of Eq.~\eqref{eq:Lambda-rho-F}
with the exponential prior on $\rho$ results in
\par
\begin{equation}
  \label{eq:Lambda-F-exp}
  \begin{split}
    \Lambda\exppr(x; \cF, \rho_0) &= \int_{0}^{\infty} d\rho \, p\exppr(\rho) \Lambda(x; \rho, \cF) \\
    &= \frac{ 2 }{ \sqrt{ 4 + \rho_0^2 } } \left[ I_0 \left( \frac{ \rho_0^2 \cF }{ 8 + 2 \rho_0^2 } \right) - I_1 \left( \frac{ \rho_0^2 \cF }{ 8 + 2 \rho_0^2 } \right) \right] e^{ \rho_0^2 \cF / ( 8 + 2 \rho_0^2 ) } \,.
  \end{split}
\end{equation}
\par\noindent
This is a strictly increasing function of $\cF$ and $\rho_0$, and is plotted
alongside $\Lambda\unifpr(x; \cF)$ in Fig.~\ref{fig:exp-prior-Lambda}.

As a third example choice of prior on $\rho$, consider the function
\par
\begin{equation}
  \label{eq:prior-rho-peak}
  p\peakpr(\rho) = \frac{ \rho }{ \rho_c^2 } e^{ -( \rho / \rho_c )^2 / 2 } \,,
\end{equation}
\par\noindent
with parameter $\rho_c$.  This function is plotted in
Fig.~\ref{fig:exp-prior-Lambda} for choices of $\rho_c$; it has a peaked shape,
with the maximum occurring at $\rho = \rho_c$.  This choice of prior is
consistent with the assumption that the signal-to-noise ratio of continuous wave
signals has some preferred value around $\rho \approx \rho_c$, as might be
expected if neutron stars possess a minimum
ellipticity~\cite{WoanEtAl2018:EvMnEllMllPl}.  Marginalisation of
Eq.~\eqref{eq:Lambda-rho-F} with this peaked prior on $\rho$ leads to
\par
\begin{equation}
  \label{eq:Lambda-F-peak}
  \begin{split}
    \Lambda\peakpr(x; \cF, \rho_c) &= \int_{0}^{\infty} d\rho \, p\peakpr(\rho) \Lambda(x; \rho, \cF) \\
    &= \frac{ 2 }{ \rho_c^2 \cF } \big[ e^{ \rho_c^2 \cF / ( 2 + \rho_c^2 ) } - 1 \big] \,.
  \end{split}
\end{equation}
\par\noindent
This is a strictly increasing function of $\cF$ and $\rho_c$, and is plotted
alongside $\Lambda\unifpr(x; \cF)$ in Fig.~\ref{fig:peak-prior-Lambda}.

\begin{figure}
\centering
\includegraphics[width=10.5 cm]{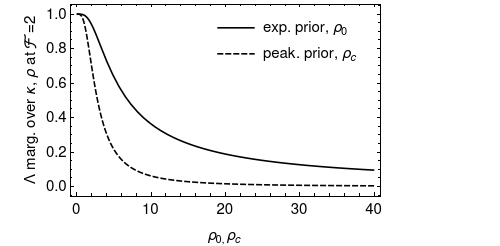}
\caption{\label{fig:Lambda-marg-kappa-rho_F-2_priors}
  The likelihood ratios marginalised over $\kappa$ and $\rho$ with the exponential
  [Eq.~\eqref{eq:Lambda-F-exp}] and peaked [Eq.~\eqref{eq:Lambda-F-peak}] priors,
  as functions of the priors' respective parameters $\rho_0$ and $\rho_c$, at
  fixed $\cF = 2$.
}
\end{figure}

Fig.~\ref{fig:Lambda-marg-kappa-rho_F-2_priors} plots the likelihood ratios
$\Lambda\exppr(x; \cF, \rho_0)$ and $\Lambda\peakpr(x; \cF, \rho_c)$
marginalised over $\kappa$ and $\rho$ with the exponential and peaked priors
respectively, as functions of the priors' respective parameters $\rho_0$ and
$\rho_c$.  The likelihoods are evaluated at $\cF = 2$, the expectation value of
$\cF$ assuming no signal is present.  The behaviour of the likelihood ratios at
$\cF = 2$ gives some indication of which hypothesis is favoured in the absence
of evidence for a signal. Both likelihood ratios favour the noise hypothesis
($\Lambda < 1$) for strictly positive parameter values.  The limiting behaviour
at zero parameter values are:
\par
\begin{align}
  \label{eq:exp-rho0-zero-limit}
  \lim_{\rho_0 \rightarrow 0} p\exppr(\rho) &= 0 \,, & \lim_{\rho_0 \rightarrow 0} \Lambda\exppr(x; \cF = 2, \rho_0) = 1 \,; \\
  \label{eq:peak-rhoc-zero-limit}
  \lim_{\rho_c \rightarrow 0} p\peakpr(\rho) &= 0 \,, & \lim_{\rho_c \rightarrow 0} \Lambda\peakpr(x; \cF = 2, \rho_c) = 1 \,.
\end{align}

\section{Receiver-operator curves}\label{sec:rece-oper-curv}

In Sec.~\ref{sec:marg-over-rho}, all three likelihood ratios marginalised over
$\rho$ [Eqs.~\eqref{eq:Lambda-F-unif},~\eqref{eq:Lambda-F-exp},
and~\eqref{eq:Lambda-F-peak}] were found to be strictly increasing functions of
$\cF$.  This implies that each marginalised likelihood ratio will have the same
detection power as the $\cF$-statistic.

Detection power is most commonly determined by Monte Carlo simulations of the
detection statistic (e.g.\ $\cF$), in both the absence and presence of a signal.
First, a set of random values of $\cF$ is generated, assuming no signal is
present.  A threshold $\cF\uthr$ is determined that gives a chosen false alarm
probability, $p\ufa$: the fraction of simulated trials where
$\cF|\unosig < \cF\uthr$.  Then, a second set of random values of $\cF$ is
generated, this time assuming the presence of a signal.  Finally, the detection
probability $p\udet$ is determined: the fraction of simulated trials where
$\cF|\usig > \cF\uthr$.  The receiver-operator curve is the function
$p\udet\big( \cF\uthr(p\ufa) \big)$.  The most powerful detection statistic is
that which gives the largest $p\udet$ at a given $p\ufa$.

If $g(\cF)$ is a strictly increasing function of $\cF$, then by definition
$\cF|\unosig < \cF\uthr$ implies $g(\cF)|\unosig < g(\cF\uthr)$, and
$\cF|\usig > \cF\uthr$ implies $g(\cF)|\usig > g(\cF\uthr)$.  Hence, by applying
$g(\cdot)$ to all simulated values of $\cF$, the transformed threshold $g(\cF\uthr)$
will yield the same false alarm and detection probabilities, and therefore
$g(\cF)$ will have the same detection power as $\cF$.

To confirm, receiver-operator curves are computed for the $\cF$-statistic,
$\cB$-statistic, and the likelihood ratio $\Lambda\unifpr(x; \cF)$
marginalised over $\kappa$ and $\rho$ with the uniform prior on $\rho$.
Following~\cite{PrixKris2009:TrSCnGrvWBVMxmSt}, the elements of $\mat{\cM}$ are fixed
at $A = 0.154$, $B = 0.234$, $C = -0.0104$, and $E = 0$, and four
signal populations are chosen:
\begin{enumerate}[label=\roman*)]
\item fixed $\rho = 4$, $\cos\iota = 0$ (i.e.\ the neutron star is viewed ``edge-on''), $\psi = 0$;
\item fixed $\rho = 4$, $\cos\iota = 0.99$~\footnote{This choice of $\cos\iota$ follows that of~\cite{PrixKris2009:TrSCnGrvWBVMxmSt}.} (i.e.\ the neutron star is viewed ``face-on''), $\psi = 0$;
\item fixed $\rho = 4$, randomly drawn $\cos\iota \in [-1, 1]$, $\psi \in [-\pi/4, \pi/4]$;
\item fixed $h_0 \sqrt{ T / \cS } = 10$, randomly drawn $\cos\iota \in [-1, 1]$, $\psi \in [-\pi/4, \pi/4]$;
\end{enumerate}
where $T = 25$~hours. For all signal populations, $\phi_0$ was randomly chosen
from $[0, 2\pi]$.  For the no-signal population, and for each of the signal
populations, $10^5$ random values of $\cF$ and $\cB$ were generated using the
program~\texttt{lalapps\_\-synthesize\-BstatMC} from the software package
LALSuite~\cite{lalsuite}; values of $\Lambda\unifpr(x; \cF)$ were then computed
from $\cF$ using Eq.~\eqref{eq:Lambda-F-unif}.

\end{paracol}

\begin{figure}[t]
\widefigure
\centering
\includegraphics[width=7.45 cm]{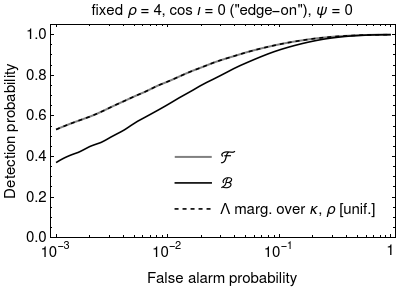}   
\includegraphics[width=7.45 cm]{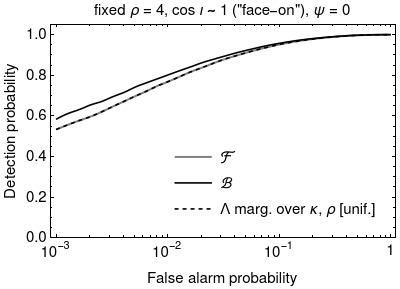}   
\vspace{6pt}
\includegraphics[width=7.45 cm]{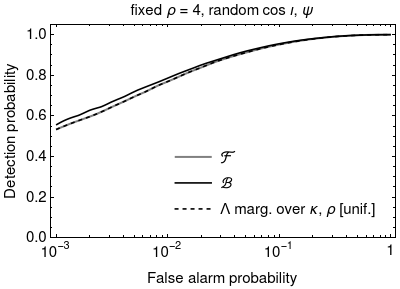}   
\includegraphics[width=7.45 cm]{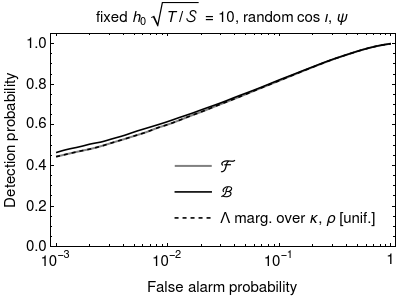}   
\caption{\label{fig:ROC}
  Receiver-operator curves for the $\cF$-statistic, $\cB$-statistic, and the
  likelihood ratio marginalised over $\kappa$ and $\rho$ with the uniform prior on
  $\rho$ [Eq.~\eqref{eq:Lambda-F-unif}], for four signal populations
  (see text).  The curves for $\Lambda\unifpr(x; \cF)$ overlay those for $\cF$.
}
\end{figure}

\begin{figure}[t]
\widefigure
\centering
\includegraphics[width=7.45 cm]{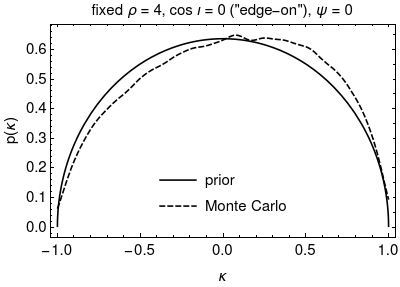}   
\includegraphics[width=7.45 cm]{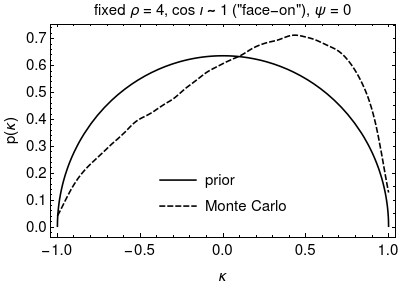}   
\vspace{6pt}
\includegraphics[width=7.45 cm]{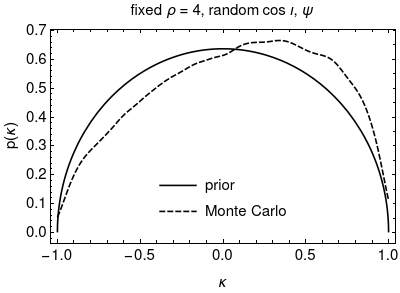}   
\includegraphics[width=7.45 cm]{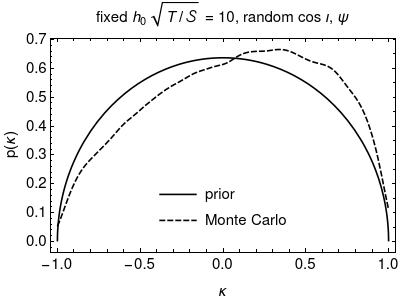}   
\caption{\label{fig:kappa-hist}
  Distribution of $\kappa$ computed from the Monte-Carlo samples compared to the
  assumed prior [Eq.~\eqref{eq:prior-kappa}], for four signal
  populations (see text).  For plotting purposes, the Monte-Carlo curves are
  smoothed with a Gaussian kernel.
}
\end{figure}

\begin{paracol}{2}
\switchcolumn

Fig.~\ref{fig:ROC} shows receiver-operator curves for the four signal
populations listed above.  The curves for the $\cF$-statistic and
$\cB$-statistic reproduce Figs.~2 and~3 of~\cite{PrixKris2009:TrSCnGrvWBVMxmSt}.
The curves for $\Lambda\unifpr(x; \cF)$ overlay the corresponding curves for
$\cF$, confirming that $\Lambda\unifpr(x; \cF)$ has identical detection power to
the $\cF$-statistic.  Receiver-operator curves for
$\Lambda\exppr(x; \cF, \rho_0)$ and $\Lambda\peakpr(x; \cF, \rho_c)$ were
computed, for various choices of $\rho_0$ and $\rho_c$ respectively, and found
to be identical to the curve for $\Lambda\unifpr(x; \cF)$.

Fig.~\ref{fig:kappa-hist} compares the distribution of $\kappa$ computed from
the Monte-Carlo samples using Eq.~\eqref{eq:kappa} with the assumed prior of
Eq.~\eqref{eq:prior-kappa}.  The Monte-Carlo distribution is a good fit to the
prior for $\cos\iota = 0$, and a poor fit for $\cos\iota \sim 1$; for the two
signal populations where $\cos\iota$ was randomly drawn, the fit is intermediate
between the two extremes.  This suggests that the initial choice of prior on
$\kappa$, which assumed no preferred orientation between the vectors $\vec{b}$
and $\vec{y}(x; p)$, is biased in favour of linearly polarised signals.  This is
consistent with $\Lambda\unifpr(x; \cF)$ being of equivalent detection power to
the $\cF$-statistic, which as noted in~\cite{PrixKris2009:TrSCnGrvWBVMxmSt} is
also biased in favour of linearly polarised signals.

\section{Discussion}\label{sec:discussion}

This paper presents an alternative approach (cf.~\cite{Derg2012:LsChrSrSWllSg,
  WhelEtAl2014:NCrdAmPrSCnGrvW, DhurEtAl2017:MrgLkFnMGrvWSr,
  BeroWhel2019:AAppBDtStCnGrvW}) to analytically marginalising the likelihood
ratio used in continuous wave searches.  Marginalised likelihood ratios were
derived assuming a prior on $\kappa$, and for example priors on $\rho$.  The
expressions for the marginalised likelihood ratios are in analytic form,
involving only exponential and Bessel functions.  Received-operator curves show
that the marginalised likelihood ratios have the same detection power as the
$\cF$-statistic, being strictly increasing functions of~$\cF$.

The marginalised likelihood ratios fail to capture the additional detection
power of the $\cB$-statistic for signal populations with randomly drawn
$\cos\iota$.  That said, as shown in~\cite{PrixKris2009:TrSCnGrvWBVMxmSt} and
reproduced in Fig.~\ref{fig:ROC}, the advantage of the $\cB$-statistic over the
$\cF$-statistic appears to be slight.  A slightly higher detection probability,
from using using the $\cB$-statistic instead of the $\cF$-statistic, corresponds
to slightly smaller $h_0$ at which continuous waves can be detected at a given
confidence $1 - p\udet$.  This small difference, however, could well be relatively
insignificant; for example, it could be within the error in $h_0$ due to the
calibration uncertainty of gravitational wave
detectors~\cite{SunEtAl2020:ChrSyErAdLClb}.  Computationally efficient
implementations of detection statistics, as exist for the $\cF$-statistic, are
essential for wide-parameter-space, computationally-costly searches.  To date,
the advantage of the $\cB$-statistic in terms of detection power have not
outweighed its disadvantage in terms of computational efficiency, and no
wide-parameter-space search for continuous wave has been performed by computing
the $\cB$-statistic directly.

It should also be noted that the $\cB$-statistic, as presented
in~\cite{PrixKris2009:TrSCnGrvWBVMxmSt}, assumes a particular emission model for
continuous waves: the triaxial model, where the neutron star radiates at twice
its rotation frequency, and the amplitudes of the ``plus'' and ``cross''
polarisations are given by $h_0 ( 1 + \cos^2\iota ) / 2$ and $h_0 \cos\iota$
respectively.  In the absence of a continuous wave detection, however, one
cannot be certain whether this is the correct emission model.  Continuous waves
radiation at other frequencies~\cite{ZimmSzed1979:GrvWRtPrRBSMAppPl,
  OwenEtAl1998:GrvWvHYRpRttNtS, VanD2005:GrvWSpNnxFPrNtS} are modelled by
different expressions for the ``plus'' and ``cross'' polarisations in terms of
$h_0$, $\cos\iota$, and other parameters.

It is possible that the detection power of the marginalised likelihood ratios
could be improved by a different choice of prior on $\kappa$.  As seen in
Fig.~\ref{fig:kappa-hist}, the prior initially assumes in
Sec.~\ref{sec:marg-over-kappa} is not necessarily a good fit, depending on the
distribution of $\cos\iota$.  If a simple analytic expression for the
distribution of $\kappa$ computed from the Monte Carlo samples could be
determined -- either from first principles, or simply as an empirical fit -- it
is possible that the likelihood ratio marginalised over $\kappa$ could still be
expressed analytically.

In marginalising the likelihood ratio over the parameters $\kappa$ and $\rho$,
it was assumed that the priors on these variables, $p(\kappa)$ and $p(\rho)$,
are independent.  Fig.~\ref{fig:kappa-hist} however shows that the prior on
$\kappa$ should be a function of $\cos\iota$, and since $\rho$ also depends on
$\cos\iota$, a joint prior $p(\kappa,\rho)$ might be needed in order to increase
detection power beyond the $\cF$-statistic.  It is unclear, however, whether a
simple but physically motivated joint prior could be found that still permits
analytic marginalisation of the likelihood ratio.  A joint prior would also make
it more difficult to change the prior on $\rho$, should one wish to consider
different models for the population of continuous wave signals.

Even if the parameterisation of the likelihood ratio in terms of $\kappa$,
$\rho$, and $\cF$ does not prove a fruitful route to obtaining an analytic
expression for the $\cB$-statistic, it could nevertheless provide a useful way
of incorporating the $\cF$-statistic into a Bayesian framework.  The
marginalised likelihood ratios presented in Sec.~\ref{sec:analyt-marg-likel} are
readily computed, being a function only of the $\cF$-statistic and well-known
special functions.  These are able to harness the computational efficiency of
existing implementations of the $\cF$-statistic, while permitting an assumption
of a prior on $\rho$ that is more physically reasonable than the prior implicit
in the $\cF$-statistic, which is biased towards stronger
signals~\cite{PrixKris2009:TrSCnGrvWBVMxmSt}.  More physically reasonable priors
on $\rho$ than the examples explored here, such as the Fermi-Dirac prior
of~\cite{PitkEtAl2017:NSmCTrSrCntGrvWP}, could be amenable to analytic
marginalisation through this approach.

An example of where a Bayesian treatment of the $\cF$-statistic could be
interesting is inferring properties of the population of Galactic neutron stars.
While methods have been proposed for inferring properties from an ensemble of
known pulsars~\cite{CutlSchu2005:GnrFsMlDtMGrvWP,
  PitkEtAl2018:HrrByMDtCnGrvWEP}, a similar framework does not yet exist for
wide-parameter-space searches.  Traditionally, such searches have computed an
\emph{upper limit} on $h_0$ satisfying the following property: given a false
alarm probability (typically 1\%, taking into account the trials factor of the
search), and assuming a population of signals with constant $h_0$ (and other
parameters chosen at random from physical priors), a high fraction of the signal
population (typically 90\%--95\%) would have been detected.  It is not expected,
however, that the population of Galactic neutron stars are all radiating
gravitational waves at the same $h_0$, and it is not clear what may be inferred
from the upper limit on $h_0$.

Perhaps, instead, a framework could be developed to compute posteriors on
parameters of an assumed model for the distribution of $h_0$; for example,
assuming the exponential prior of $\rho$ of Eq.~\eqref{eq:prior-rho-exp}, and
inferring the posterior on its parameter $\rho_c$ from a wide-parameter-space
search.  The approach to marginalisation of the likelihood ratio presented in
this paper might provide a route towards constructing this framework.

\vspace{6pt}

\funding{This research was supported by the Australian Research Council Centre
  of Excellence for Gravitational Wave Discovery (OzGrav) through project number
  CE170100004.}

\acknowledgments{The author thanks the continuous wave working group of the LIGO
  Scientific Collaboration, Virgo Collaboration, and KAGRA Collaboration for
  helpful comments.  This research used the software LALSuite~\cite{lalsuite}
  and Mathematica~\cite{Mathematica}.}

\conflictsofinterest{The author declares no conflict of interest.}

\end{paracol}

\reftitle{References}

\externalbibliography{yes}


\begin{thebibliography}{999}

\bibitem[{Aasi} \em{et~al.}(2015){Aasi} et~al.]{LIGO2015:AdvLIG}
{Aasi}, J.; others.
\newblock {Advanced LIGO}.
\newblock {\em Classical and Quantum Gravity} {\bf 2015}, {\em 32},~074001,
  \href{http://xxx.lanl.gov/abs/1411.4547}{{\normalfont
  [arXiv:gr-qc/1411.4547]}}.
\newblock
  doi:{\changeurlcolor{black}\href{https://doi.org/10.1088/0264-9381/32/7/074001}{\detokenize{10.1088/0264-9381/32/7/074001}}}.

\bibitem[{Acernese} \em{et~al.}(2018){Acernese} et~al.]{AcerEtAl2018:SttAdvVr}
{Acernese}, F.; others.
\newblock {Status of Advanced Virgo}.
\newblock  {European Physical Journal Web of Conferences},  2018, Vol. 182, p.
  02003.
\newblock
  doi:{\changeurlcolor{black}\href{https://doi.org/10.1051/epjconf/201818202003}{\detokenize{10.1051/epjconf/201818202003}}}.

\bibitem[{Abbott} \em{et~al.}(2020){Abbott}
  et~al.]{KAGRLIGOVirg2020:PrObLclGrvTrALAVK}
{Abbott}, B.P.; others.
\newblock {Prospects for observing and localizing gravitational-wave transients
  with Advanced LIGO, Advanced Virgo and KAGRA}.
\newblock {\em Living Reviews in Relativity} {\bf 2020}, {\em 23},~3.
\newblock
  doi:{\changeurlcolor{black}\href{https://doi.org/10.1007/s41114-020-00026-9}{\detokenize{10.1007/s41114-020-00026-9}}}.

\bibitem[{Abbott} \em{et~al.}(2019{\natexlab{a}}){Abbott}
  et~al.]{LIGOVirg2019:ASCntGrvWIsNtSUAdLOD}
{Abbott}, B.P.; others.
\newblock {All-sky search for continuous gravitational waves from isolated
  neutron stars using Advanced LIGO O2 data}.
\newblock {\em Physical Review D} {\bf 2019}, {\em 100},~024004,
  \href{http://xxx.lanl.gov/abs/1903.01901}{{\normalfont
  [arXiv:astro-ph.HE/1903.01901]}}.
\newblock
  doi:{\changeurlcolor{black}\href{https://doi.org/10.1103/PhysRevD.100.024004}{\detokenize{10.1103/PhysRevD.100.024004}}}.

\bibitem[{Abbott} \em{et~al.}(2019{\natexlab{b}}){Abbott}
  et~al.]{LIGOVirg2019:NrrSrGrvWKPlUSLObR}
{Abbott}, B.P.; others.
\newblock {Narrow-band search for gravitational waves from known pulsars using
  the second LIGO observing run}.
\newblock {\em Physical Review D} {\bf 2019}, {\em 99},~122002.
\newblock
  doi:{\changeurlcolor{black}\href{https://doi.org/10.1103/PhysRevD.99.122002}{\detokenize{10.1103/PhysRevD.99.122002}}}.

\bibitem[{Abbott} \em{et~al.}(2019{\natexlab{c}}){Abbott}
  et~al.]{LIGOVirg2019:SGrvWScXSAdLObsRImHMM}
{Abbott}, B.P.; others.
\newblock {Search for gravitational waves from Scorpius X-1 in the second
  Advanced LIGO observing run with an improved hidden Markov model}.
\newblock {\em Physical Review D} {\bf 2019}, {\em 100},~122002.
\newblock
  doi:{\changeurlcolor{black}\href{https://doi.org/10.1103/PhysRevD.100.122002}{\detokenize{10.1103/PhysRevD.100.122002}}}.

\bibitem[{Abbott} \em{et~al.}(2019{\natexlab{d}}){Abbott}
  et~al.]{LIGOVirg2019:SrGrvWvKPlTHrm20152017LD}
{Abbott}, B.P.; others.
\newblock {Searches for Gravitational Waves from Known Pulsars at Two Harmonics
  in 2015-2017 LIGO Data}.
\newblock {\em Astrophysical Journal} {\bf 2019}, {\em 879},~10,
  \href{http://xxx.lanl.gov/abs/1902.08507}{{\normalfont
  [arXiv:astro-ph.HE/1902.08507]}}.
\newblock
  doi:{\changeurlcolor{black}\href{https://doi.org/10.3847/1538-4357/ab20cb}{\detokenize{10.3847/1538-4357/ab20cb}}}.

\bibitem[{Palomba} \em{et~al.}(2019){Palomba}
  et~al.]{PaloEtAl2019:DCnsUlBMSrCnGrvW}
{Palomba}, C.; others.
\newblock {Direct Constraints on the Ultralight Boson Mass from Searches of
  Continuous Gravitational Waves}.
\newblock {\em Physical Review Letters} {\bf 2019}, {\em 123},~171101,
  \href{http://xxx.lanl.gov/abs/1909.08854}{{\normalfont
  [arXiv:astro-ph.HE/1909.08854]}}.
\newblock
  doi:{\changeurlcolor{black}\href{https://doi.org/10.1103/PhysRevLett.123.171101}{\detokenize{10.1103/PhysRevLett.123.171101}}}.

\bibitem[{Covas} and {Sintes}(2020)]{CovaSint2020:FASCntGrvSgUnNSBSUAdLD}
{Covas}, P.B.; {Sintes}, A.M.
\newblock {First All-Sky Search for Continuous Gravitational-Wave Signals from
  Unknown Neutron Stars in Binary Systems Using Advanced LIGO Data}.
\newblock {\em Physical Review Letters} {\bf 2020}, {\em 124},~191102,
  \href{http://xxx.lanl.gov/abs/2001.08411}{{\normalfont
  [arXiv:gr-qc/2001.08411]}}.
\newblock
  doi:{\changeurlcolor{black}\href{https://doi.org/10.1103/PhysRevLett.124.191102}{\detokenize{10.1103/PhysRevLett.124.191102}}}.

\bibitem[{Dergachev} and {Papa}(2020)]{DergPapa2020:RsFASCnGrvWSmlSr}
{Dergachev}, V.; {Papa}, M.A.
\newblock {Results from the First All-Sky Search for Continuous Gravitational
  Waves from Small-Ellipticity Sources}.
\newblock {\em Physical Review Letters} {\bf 2020}, {\em 125},~171101,
  \href{http://xxx.lanl.gov/abs/2004.08334}{{\normalfont
  [arXiv:gr-qc/2004.08334]}}.
\newblock
  doi:{\changeurlcolor{black}\href{https://doi.org/10.1103/PhysRevLett.125.171101}{\detokenize{10.1103/PhysRevLett.125.171101}}}.

\bibitem[{Fesik} and {Papa}(2020)]{FesiPapa2020:FrSrRmGrvWPJ05}
{Fesik}, L.; {Papa}, M.A.
\newblock {First Search for r-mode Gravitational Waves from PSR J0537-6910}.
\newblock {\em Astrophysical Journal} {\bf 2020}, {\em 895},~11,
  \href{http://xxx.lanl.gov/abs/2001.07605}{{\normalfont
  [arXiv:gr-qc/2001.07605]}}.
\newblock
  doi:{\changeurlcolor{black}\href{https://doi.org/10.3847/1538-4357/ab8193}{\detokenize{10.3847/1538-4357/ab8193}}}.

\bibitem[{Lindblom} and {Owen}(2020)]{LindOwen2020:DrSCntGrvWTSpRDALSObR}
{Lindblom}, L.; {Owen}, B.J.
\newblock {Directed searches for continuous gravitational waves from twelve
  supernova remnants in data from Advanced LIGO's second observing run}.
\newblock {\em Physical Review D} {\bf 2020}, {\em 101},~083023,
  \href{http://xxx.lanl.gov/abs/2003.00072}{{\normalfont
  [arXiv:gr-qc/2003.00072]}}.
\newblock
  doi:{\changeurlcolor{black}\href{https://doi.org/10.1103/PhysRevD.101.083023}{\detokenize{10.1103/PhysRevD.101.083023}}}.

\bibitem[{Middleton} \em{et~al.}(2020){Middleton}, {Clearwater}, {Melatos}, and
  {Dunn}]{MiddEtAl2020:SGrvWFLMXBnSAdLObsRImHMM}
{Middleton}, H.; {Clearwater}, P.; {Melatos}, A.; {Dunn}, L.
\newblock {Search for gravitational waves from five low mass x-ray binaries in
  the second Advanced LIGO observing run with an improved hidden Markov model}.
\newblock {\em Physical Review D} {\bf 2020}, {\em 102},~023006,
  \href{http://xxx.lanl.gov/abs/2006.06907}{{\normalfont
  [arXiv:astro-ph.HE/2006.06907]}}.
\newblock
  doi:{\changeurlcolor{black}\href{https://doi.org/10.1103/PhysRevD.102.023006}{\detokenize{10.1103/PhysRevD.102.023006}}}.

\bibitem[{Millhouse} \em{et~al.}(2020){Millhouse}, {Strang}, and
  {Melatos}]{MillEtAl2020:SGrvW12YSprRmHMMAdLSObR}
{Millhouse}, M.; {Strang}, L.; {Melatos}, A.
\newblock {Search for gravitational waves from 12 young supernova remnants with
  a hidden Markov model in Advanced LIGO's second observing run}.
\newblock {\em Physical Review D} {\bf 2020}, {\em 102},~083025,
  \href{http://xxx.lanl.gov/abs/2003.08588}{{\normalfont
  [arXiv:gr-qc/2003.08588]}}.
\newblock
  doi:{\changeurlcolor{black}\href{https://doi.org/10.1103/PhysRevD.102.083025}{\detokenize{10.1103/PhysRevD.102.083025}}}.

\bibitem[{Piccinni} \em{et~al.}(2020){Piccinni}
  et~al.]{PiccEtAl2020:DrSCntGrvSGlCALSObR}
{Piccinni}, O.J.; others.
\newblock {Directed search for continuous gravitational-wave signals from the
  Galactic Center in the Advanced LIGO second observing run}.
\newblock {\em Physical Review D} {\bf 2020}, {\em 101},~082004,
  \href{http://xxx.lanl.gov/abs/1910.05097}{{\normalfont
  [arXiv:gr-qc/1910.05097]}}.
\newblock
  doi:{\changeurlcolor{black}\href{https://doi.org/10.1103/PhysRevD.101.082004}{\detokenize{10.1103/PhysRevD.101.082004}}}.

\bibitem[{Sun} \em{et~al.}(2020){Sun}, {Brito}, and
  {Isi}]{SunEtAl2020:SrUltBsCyXAdvL}
{Sun}, L.; {Brito}, R.; {Isi}, M.
\newblock {Search for ultralight bosons in Cygnus X-1 with Advanced LIGO}.
\newblock {\em Physical Review D} {\bf 2020}, {\em 101},~063020,
  \href{http://xxx.lanl.gov/abs/1909.11267}{{\normalfont
  [arXiv:gr-qc/1909.11267]}}.
\newblock
  doi:{\changeurlcolor{black}\href{https://doi.org/10.1103/PhysRevD.101.063020}{\detokenize{10.1103/PhysRevD.101.063020}}}.

\bibitem[{Zhang} \em{et~al.}(2021){Zhang}, {Papa}, {Krishnan}, and
  {Watts}]{ZhanEtAl2021:SrCntGrvWvScXLOD}
{Zhang}, Y.; {Papa}, M.A.; {Krishnan}, B.; {Watts}, A.L.
\newblock {Search for Continuous Gravitational Waves from Scorpius X-1 in LIGO
  O2 Data}.
\newblock {\em Astrophysical Journal Letters} {\bf 2021}, {\em 906},~L14,
  \href{http://xxx.lanl.gov/abs/2011.04414}{{\normalfont
  [arXiv:astro-ph.HE/2011.04414]}}.
\newblock
  doi:{\changeurlcolor{black}\href{https://doi.org/10.3847/2041-8213/abd256}{\detokenize{10.3847/2041-8213/abd256}}}.

\bibitem[{Jones} and {Sun}(2021)]{JoneSun2021:SCntGrvWFmBSAdLObRHMM}
{Jones}, D.; {Sun}, L.
\newblock {Search for continuous gravitational waves from Fomalhaut b in the
  second Advanced LIGO observing run with a hidden Markov model}.
\newblock {\em Physical Review D} {\bf 2021}, {\em 103},~023020,
  \href{http://xxx.lanl.gov/abs/2007.08732}{{\normalfont
  [arXiv:gr-qc/2007.08732]}}.
\newblock
  doi:{\changeurlcolor{black}\href{https://doi.org/10.1103/PhysRevD.103.023020}{\detokenize{10.1103/PhysRevD.103.023020}}}.

\bibitem[{Beniwal} \em{et~al.}(2021){Beniwal}, {Clearwater}, {Dunn}, {Melatos},
  and {Ottaway}]{BeniEtAl2021:SrCntGrvWTHSrUHdMM}
{Beniwal}, D.; {Clearwater}, P.; {Dunn}, L.; {Melatos}, A.; {Ottaway}, D.
\newblock {Search for continuous gravitational waves from ten H.E.S.S. sources
  using a hidden Markov model}.
\newblock {\em Physical Review D} {\bf 2021}, {\em 103},~083009.
\newblock
  doi:{\changeurlcolor{black}\href{https://doi.org/10.1103/PhysRevD.103.083009}{\detokenize{10.1103/PhysRevD.103.083009}}}.

\bibitem[{Wette} \em{et~al.}(2021){Wette}, {Dunn}, {Clearwater}, and
  {Melatos}]{1HzSearch}
{Wette}, K.; {Dunn}, L.; {Clearwater}, P.; {Melatos}, A.
\newblock {Deep exploration for continuous gravitational waves at 171--172 Hz
  in LIGO second observing run data}.
\newblock {\em Phys. Rev. D} {\bf 2021}, {\em 103},~083020.
\newblock
  doi:{\changeurlcolor{black}\href{https://doi.org/10.1103/PhysRevD.103.083020}{\detokenize{10.1103/PhysRevD.103.083020}}}.

\bibitem[{Abbott} \em{et~al.}(2020){Abbott} et~al.]{LIGOVirg2020:GrvCnsEqElMlP}
{Abbott}, R.; others.
\newblock {Gravitational-wave Constraints on the Equatorial Ellipticity of
  Millisecond Pulsars}.
\newblock {\em Astrophysical Journal} {\bf 2020}, {\em 902},~L21.
\newblock
  doi:{\changeurlcolor{black}\href{https://doi.org/10.3847/2041-8213/abb655}{\detokenize{10.3847/2041-8213/abb655}}}.

\bibitem[{Abbott} \em{et~al.}(2021){Abbott}
  et~al.]{LIGOVirg2021:ASEOLDCntGrvSgUnNtSBS}
{Abbott}, R.; others.
\newblock {All-sky search in early O3 LIGO data for continuous
  gravitational-wave signals from unknown neutron stars in binary systems}.
\newblock {\em Physical Review D} {\bf 2021}, {\em 103},~064017,
  \href{http://xxx.lanl.gov/abs/2012.12128}{{\normalfont
  [arXiv:gr-qc/2012.12128]}}.
\newblock
  doi:{\changeurlcolor{black}\href{https://doi.org/10.1103/PhysRevD.103.064017}{\detokenize{10.1103/PhysRevD.103.064017}}}.

\bibitem[{Bonazzola} and {Gourgoulhon}(1996)]{BonaGour1996:GrvWPlEmMgFIDst}
{Bonazzola}, S.; {Gourgoulhon}, E.
\newblock {Gravitational waves from pulsars: emission by the magnetic field
  induced distortion}.
\newblock {\em Astronomy \& Astrophysics} {\bf 1996}, {\em 312},~675,
  \href{http://xxx.lanl.gov/abs/astro-ph/9602107}{{\normalfont
  [astro-ph/9602107]}}.

\bibitem[{Ushomirsky} \em{et~al.}(2000){Ushomirsky}, {Cutler}, and
  {Bildsten}]{UshoEtAl2000:DfrAcNtSCGrvWEm}
{Ushomirsky}, G.; {Cutler}, C.; {Bildsten}, L.
\newblock {Deformations of accreting neutron star crusts and gravitational wave
  emission}.
\newblock {\em Monthly Notices of the Royal Astronomical Society} {\bf 2000},
  {\em 319},~902,  \href{http://xxx.lanl.gov/abs/astro-ph/0001136}{{\normalfont
  [astro-ph/0001136]}}.
\newblock
  doi:{\changeurlcolor{black}\href{https://doi.org/10.1046/j.1365-8711.2000.03938.x}{\detokenize{10.1046/j.1365-8711.2000.03938.x}}}.

\bibitem[{Owen}(2005)]{Owen2005:MxElDfrCmSEEqtS}
{Owen}, B.J.
\newblock {Maximum Elastic Deformations of Compact Stars with Exotic Equations
  of State}.
\newblock {\em Physical Review Letters} {\bf 2005}, {\em 95},~211101,
  \href{http://xxx.lanl.gov/abs/astro-ph/0503399}{{\normalfont
  [astro-ph/0503399]}}.
\newblock
  doi:{\changeurlcolor{black}\href{https://doi.org/10.1103/PhysRevLett.95.211101}{\detokenize{10.1103/PhysRevLett.95.211101}}}.

\bibitem[{Haskell} \em{et~al.}(2008){Haskell}, {Samuelsson}, {Glampedakis}, and
  {Andersson}]{HaskEtAl2008:MdlMgnDfNtSt}
{Haskell}, B.; {Samuelsson}, L.; {Glampedakis}, K.; {Andersson}, N.
\newblock {Modelling magnetically deformed neutron stars}.
\newblock {\em Monthly Notices of the Royal Astronomical Society} {\bf 2008},
  {\em 385},~531,  \href{http://xxx.lanl.gov/abs/0705.1780}{{\normalfont
  [arXiv:astro-ph/0705.1780]}}.
\newblock
  doi:{\changeurlcolor{black}\href{https://doi.org/10.1111/j.1365-2966.2008.12861.x}{\detokenize{10.1111/j.1365-2966.2008.12861.x}}}.

\bibitem[{Glampedakis} \em{et~al.}(2012){Glampedakis}, {Jones}, and
  {Samuelsson}]{GlamEtAl2012:GrvWvClrMnNtS}
{Glampedakis}, K.; {Jones}, D.I.; {Samuelsson}, L.
\newblock {Gravitational Waves from Color-Magnetic ``Mountains'' in Neutron
  Stars}.
\newblock {\em Physical Review Letters} {\bf 2012}, {\em 109},~081103,
  \href{http://xxx.lanl.gov/abs/1204.3781}{{\normalfont
  [arXiv:astro-ph.SR/1204.3781]}}.
\newblock
  doi:{\changeurlcolor{black}\href{https://doi.org/10.1103/PhysRevLett.109.081103}{\detokenize{10.1103/PhysRevLett.109.081103}}}.

\bibitem[{Johnson-McDaniel} and {Owen}(2013)]{JohnOwen2013:MxElDfrRltSt}
{Johnson-McDaniel}, N.K.; {Owen}, B.J.
\newblock {Maximum elastic deformations of relativistic stars}.
\newblock {\em Physical Review D} {\bf 2013}, {\em 88},~044004,
  \href{http://xxx.lanl.gov/abs/1208.5227}{{\normalfont
  [arXiv:astro-ph.SR/1208.5227]}}.
\newblock
  doi:{\changeurlcolor{black}\href{https://doi.org/10.1103/PhysRevD.88.044004}{\detokenize{10.1103/PhysRevD.88.044004}}}.

\bibitem[{Woan} \em{et~al.}(2018){Woan}, {Pitkin}, {Haskell}, {Jones}, and
  {Lasky}]{WoanEtAl2018:EvMnEllMllPl}
{Woan}, G.; {Pitkin}, M.D.; {Haskell}, B.; {Jones}, D.I.; {Lasky}, P.D.
\newblock {Evidence for a Minimum Ellipticity in Millisecond Pulsars}.
\newblock {\em Astrophysical Journal} {\bf 2018}, {\em 863},~L40,
  \href{http://xxx.lanl.gov/abs/1806.02822}{{\normalfont
  [arXiv:astro-ph.HE/1806.02822]}}.
\newblock
  doi:{\changeurlcolor{black}\href{https://doi.org/10.3847/2041-8213/aad86a}{\detokenize{10.3847/2041-8213/aad86a}}}.

\bibitem[{Osborne} and {Jones}(2020)]{OsboJone2020:GrvWMgnInThNSMn}
{Osborne}, E.L.; {Jones}, D.I.
\newblock {Gravitational waves from magnetically induced thermal neutron star
  mountains}.
\newblock {\em Monthly Notices of the Royal Astronomical Society} {\bf 2020},
  {\em 494},~2839--2850,
  \href{http://xxx.lanl.gov/abs/1910.04453}{{\normalfont
  [arXiv:astro-ph.HE/1910.04453]}}.
\newblock
  doi:{\changeurlcolor{black}\href{https://doi.org/10.1093/mnras/staa858}{\detokenize{10.1093/mnras/staa858}}}.

\bibitem[{Neyman} and {Pearson}(1933)]{NeymPear1933:PrMEfTsSttHyp}
{Neyman}, J.; {Pearson}, E.S.
\newblock {On the Problem of the Most Efficient Tests of Statistical
  Hypotheses}.
\newblock {\em Philosophical Transactions of the Royal Society A} {\bf 1933},
  {\em 231},~289.
\newblock
  doi:{\changeurlcolor{black}\href{https://doi.org/10.1098/rsta.1933.0009}{\detokenize{10.1098/rsta.1933.0009}}}.

\bibitem[{Jaranowski} \em{et~al.}(1998){Jaranowski}, {Kr{\'o}lak}, and
  {Schutz}]{JaraEtAl1998:DAnGrvSgSpNSSDtc}
{Jaranowski}, P.; {Kr{\'o}lak}, A.; {Schutz}, B.F.
\newblock {Data analysis of gravitational-wave signals from spinning neutron
  stars: The signal and its detection}.
\newblock {\em Physical Review D} {\bf 1998}, {\em 58},~063001,
  \href{http://xxx.lanl.gov/abs/gr-qc/9804014}{{\normalfont [gr-qc/9804014]}}.
\newblock
  doi:{\changeurlcolor{black}\href{https://doi.org/10.1103/PhysRevD.58.063001}{\detokenize{10.1103/PhysRevD.58.063001}}}.

\bibitem[{Thrane} and {Talbot}(2019)]{ThraTalb2019:IntBInGrvAPEsMSHrM}
{Thrane}, E.; {Talbot}, C.
\newblock {An introduction to Bayesian inference in gravitational-wave
  astronomy: Parameter estimation, model selection, and hierarchical models}.
\newblock {\em Publications of the Astronomical Society of Australia} {\bf
  2019}, {\em 36},~e010,
  \href{http://xxx.lanl.gov/abs/1809.02293}{{\normalfont
  [arXiv:astro-ph.IM/1809.02293]}}.
\newblock
  doi:{\changeurlcolor{black}\href{https://doi.org/10.1017/pasa.2019.2}{\detokenize{10.1017/pasa.2019.2}}}.

\bibitem[{Searle} \em{et~al.}(2008){Searle}, {Sutton}, {Tinto}, and
  {Woan}]{SearEtAl2008:RbByDtcUnmBr}
{Searle}, A.C.; {Sutton}, P.J.; {Tinto}, M.; {Woan}, G.
\newblock {Robust Bayesian detection of unmodelled bursts}.
\newblock {\em Classical and Quantum Gravity} {\bf 2008}, {\em 25},~114038,
  \href{http://xxx.lanl.gov/abs/0712.0196}{{\normalfont
  [arXiv:gr-qc/0712.0196]}}.
\newblock
  doi:{\changeurlcolor{black}\href{https://doi.org/10.1088/0264-9381/25/11/114038}{\detokenize{10.1088/0264-9381/25/11/114038}}}.

\bibitem[{Searle}(2008)]{Sear2008:MntByTcGrvWBDAn}
{Searle}, A.C.
\newblock {Monte-Carlo and Bayesian techniques in gravitational wave burst data
  analysis}.
\newblock {\em arXiv} {\bf 2008},
  \href{http://xxx.lanl.gov/abs/0804.1161}{{\normalfont
  [arXiv:gr-qc/0804.1161]}}.

\bibitem[{Prix} and {Krishnan}(2009)]{PrixKris2009:TrSCnGrvWBVMxmSt}
{Prix}, R.; {Krishnan}, B.
\newblock {Targeted search for continuous gravitational waves: Bayesian versus
  maximum-likelihood statistics}.
\newblock {\em Classical and Quantum Gravity} {\bf 2009}, {\em 26},~204013,
  \href{http://xxx.lanl.gov/abs/0907.2569}{{\normalfont
  [arXiv:gr-qc/0907.2569]}}.
\newblock
  doi:{\changeurlcolor{black}\href{https://doi.org/10.1088/0264-9381/26/20/204013}{\detokenize{10.1088/0264-9381/26/20/204013}}}.

\bibitem[{Prix}(2010)]{Prix2010:FsImpCmp}
{Prix}, R.
\newblock {The $\mathcal{F}$-statistic and its implementation in
  ComputeFStatistic\_v2}.
\newblock Technical Report T0900149-v5, LIGO,  2010.

\bibitem[{Patel} \em{et~al.}(2010){Patel}, {Siemens}, {Dupuis}, and
  {Betzwieser}]{PateEtAl2010:ImpBrRsCnWSGrvWD}
{Patel}, P.; {Siemens}, X.; {Dupuis}, R.; {Betzwieser}, J.
\newblock {Implementation of barycentric resampling for continuous wave
  searches in gravitational wave data}.
\newblock {\em Physical Review D} {\bf 2010}, {\em 81},~084032,
  \href{http://xxx.lanl.gov/abs/0912.4255}{{\normalfont
  [arXiv:gr-qc/0912.4255]}}.
\newblock
  doi:{\changeurlcolor{black}\href{https://doi.org/10.1103/PhysRevD.81.084032}{\detokenize{10.1103/PhysRevD.81.084032}}}.

\bibitem[{Poghosyan} \em{et~al.}(2015){Poghosyan}, {Matta}, {Streit}, {Bejger},
  and {Kr{\'o}lak}]{PoghEtAl2015:ArImpPrlSfSPGrWS}
{Poghosyan}, G.; {Matta}, S.; {Streit}, A.; {Bejger}, M.; {Kr{\'o}lak}, A.
\newblock {Architecture, implementation and parallelization of the software to
  search for periodic gravitational wave signals}.
\newblock {\em Computer Physics Communications} {\bf 2015}, {\em
  188},~167--176,  \href{http://xxx.lanl.gov/abs/1410.3677}{{\normalfont
  [arXiv:gr-qc/1410.3677]}}.
\newblock
  doi:{\changeurlcolor{black}\href{https://doi.org/10.1016/j.cpc.2014.10.025}{\detokenize{10.1016/j.cpc.2014.10.025}}}.

\bibitem[{Dergachev}(2012)]{Derg2012:LsChrSrSWllSg}
{Dergachev}, V.
\newblock {Loosely coherent searches for sets of well-modeled signals}.
\newblock {\em Physical Review D} {\bf 2012}, {\em 85},~062003,
  \href{http://xxx.lanl.gov/abs/1110.3297}{{\normalfont
  [arXiv:gr-qc/1110.3297]}}.
\newblock
  doi:{\changeurlcolor{black}\href{https://doi.org/10.1103/PhysRevD.85.062003}{\detokenize{10.1103/PhysRevD.85.062003}}}.

\bibitem[{Whelan} \em{et~al.}(2014){Whelan}, {Prix}, {Cutler}, and
  {Willis}]{WhelEtAl2014:NCrdAmPrSCnGrvW}
{Whelan}, J.T.; {Prix}, R.; {Cutler}, C.J.; {Willis}, J.L.
\newblock {New coordinates for the amplitude parameter space of continuous
  gravitational waves}.
\newblock {\em Classical and Quantum Gravity} {\bf 2014}, {\em 31},~065002,
  \href{http://xxx.lanl.gov/abs/1311.0065}{{\normalfont
  [arXiv:gr-qc/1311.0065]}}.
\newblock
  doi:{\changeurlcolor{black}\href{https://doi.org/10.1088/0264-9381/31/6/065002}{\detokenize{10.1088/0264-9381/31/6/065002}}}.

\bibitem[{Dhurandhar} \em{et~al.}(2017){Dhurandhar}, {Krishnan}, and
  {Willis}]{DhurEtAl2017:MrgLkFnMGrvWSr}
{Dhurandhar}, S.; {Krishnan}, B.; {Willis}, J.L.
\newblock {Marginalizing the likelihood function for modeled gravitational wave
  searches}.
\newblock {\em arXiv} {\bf 2017},
  \href{http://xxx.lanl.gov/abs/1707.08163}{{\normalfont [1707.08163]}}.

\bibitem[{Bero} and {Whelan}(2019)]{BeroWhel2019:AAppBDtStCnGrvW}
{Bero}, J.J.; {Whelan}, J.T.
\newblock {An analytic approximation to the Bayesian detection statistic for
  continuous gravitational waves}.
\newblock {\em Classical and Quantum Gravity} {\bf 2019}, {\em 36},~015013,
  \href{http://xxx.lanl.gov/abs/1808.05453}{{\normalfont
  [arXiv:gr-qc/1808.05453]}}.
\newblock
  doi:{\changeurlcolor{black}\href{https://doi.org/10.1088/1361-6382/aaed6a}{\detokenize{10.1088/1361-6382/aaed6a}}}.

\bibitem[{Prix}(2007)]{Prix2007:SrCnGrvWMMltFs}
{Prix}, R.
\newblock {Search for continuous gravitational waves: Metric of the
  multidetector $\mathcal{F}$-statistic}.
\newblock {\em Physical Review D} {\bf 2007}, {\em 75},~023004,
  \href{http://xxx.lanl.gov/abs/gr-qc/0606088}{{\normalfont [gr-qc/0606088]}}.
\newblock
  doi:{\changeurlcolor{black}\href{https://doi.org/10.1103/PhysRevD.75.023004}{\detokenize{10.1103/PhysRevD.75.023004}}}.

\bibitem[{Brady} \em{et~al.}(1998){Brady}, {Creighton}, {Cutler}, and
  {Schutz}]{BradEtAl1998:SrcPrdSrLI}
{Brady}, P.R.; {Creighton}, T.; {Cutler}, C.; {Schutz}, B.F.
\newblock {Searching for periodic sources with LIGO}.
\newblock {\em Physical Review D} {\bf 1998}, {\em 57},~2101,
  \href{http://xxx.lanl.gov/abs/gr-qc/9702050}{{\normalfont [gr-qc/9702050]}}.
\newblock
  doi:{\changeurlcolor{black}\href{https://doi.org/10.1103/PhysRevD.57.2101}{\detokenize{10.1103/PhysRevD.57.2101}}}.

\bibitem[{Balasubramanian} \em{et~al.}(1996){Balasubramanian}, {Sathyaprakash},
  and {Dhurandhar}]{BalaEtAl1996:GrvWClsBDStMCEsPr}
{Balasubramanian}, R.; {Sathyaprakash}, B.S.; {Dhurandhar}, S.V.
\newblock {Gravitational waves from coalescing binaries: Detection strategies
  and Monte Carlo estimation of parameters}.
\newblock {\em Physical Review D} {\bf 1996}, {\em 53},~3033,
  \href{http://xxx.lanl.gov/abs/gr-qc/9508011}{{\normalfont [gr-qc/9508011]}}.
\newblock
  doi:{\changeurlcolor{black}\href{https://doi.org/10.1103/PhysRevD.53.3033}{\detokenize{10.1103/PhysRevD.53.3033}}}.

\bibitem[{Owen}(1996)]{Owen1996:STmGrvWInsBnCTmS}
{Owen}, B.J.
\newblock {Search templates for gravitational waves from inspiraling binaries:
  Choice of template spacing}.
\newblock {\em Physical Review D} {\bf 1996}, {\em 53},~6749,
  \href{http://xxx.lanl.gov/abs/gr-qc/9511032}{{\normalfont [gr-qc/9511032]}}.
\newblock
  doi:{\changeurlcolor{black}\href{https://doi.org/10.1103/PhysRevD.53.6749}{\detokenize{10.1103/PhysRevD.53.6749}}}.

\bibitem[{Wette} and {Prix}(2013)]{WettPrix2013:FPrmMtASrGrvPl}
{Wette}, K.; {Prix}, R.
\newblock {Flat parameter-space metric for all-sky searches for
  gravitational-wave pulsars}.
\newblock {\em Physical Review D} {\bf 2013}, {\em 88},~123005,
  \href{http://xxx.lanl.gov/abs/1310.5587}{{\normalfont
  [arXiv:gr-qc/1310.5587]}}.
\newblock
  doi:{\changeurlcolor{black}\href{https://doi.org/10.1103/PhysRevD.88.123005}{\detokenize{10.1103/PhysRevD.88.123005}}}.

\bibitem[{Kr{\'o}lak} \em{et~al.}(2004){Kr{\'o}lak}, {Tinto}, and
  {Vallisneri}]{KrolEtAl2004:OptFltLIDt}
{Kr{\'o}lak}, A.; {Tinto}, M.; {Vallisneri}, M.
\newblock {Optimal filtering of the LISA data}.
\newblock {\em Physical Review D} {\bf 2004}, {\em 70},~022003,
  \href{http://xxx.lanl.gov/abs/gr-qc/0401108}{{\normalfont [gr-qc/0401108]}}.
\newblock
  doi:{\changeurlcolor{black}\href{https://doi.org/10.1103/PhysRevD.70.022003}{\detokenize{10.1103/PhysRevD.70.022003}}}.

\bibitem[{Whelan} \em{et~al.}(2008){Whelan}, {Prix}, and
  {Khurana}]{WhelEtAl2008:ImSGlWhtBMLDCh1UFstTB}
{Whelan}, J.T.; {Prix}, R.; {Khurana}, D.
\newblock {Improved search for galactic white-dwarf binaries in Mock LISA Data
  Challenge 1B using an $\mathcal{F}$-statistic template bank}.
\newblock {\em Classical and Quantum Gravity} {\bf 2008}, {\em 25},~184029,
  \href{http://xxx.lanl.gov/abs/0805.1972}{{\normalfont
  [arXiv:gr-qc/0805.1972]}}.
\newblock
  doi:{\changeurlcolor{black}\href{https://doi.org/10.1088/0264-9381/25/18/184029}{\detokenize{10.1088/0264-9381/25/18/184029}}}.

\bibitem[{Marsaglia}(1972)]{Mars1972:ChsPnSrfSp}
{Marsaglia}, G.
\newblock {Choosing a Point from the Surface of a Sphere}.
\newblock {\em Annals of Mathematical Statistics} {\bf 1972}, {\em 43},~645.
\newblock
  doi:{\changeurlcolor{black}\href{https://doi.org/10.1214/aoms/1177692644}{\detokenize{10.1214/aoms/1177692644}}}.

\bibitem[{LIGO Scientific Collaboration}(2018)]{lalsuite}
{LIGO Scientific Collaboration}.
\newblock {LIGO} {A}lgorithm {L}ibrary - {LALS}uite.
\newblock Free software (GPL),  2018.
\newblock
  doi:{\changeurlcolor{black}\href{https://doi.org/10.7935/GT1W-FZ16}{\detokenize{10.7935/GT1W-FZ16}}}.

\bibitem[{Sun} \em{et~al.}(2020){Sun} et~al.]{SunEtAl2020:ChrSyErAdLClb}
{Sun}, L.; others.
\newblock {Characterization of systematic error in Advanced {LIGO}
  calibration}.
\newblock {\em Classical and Quantum Gravity} {\bf 2020}, {\em 37},~225008.
\newblock
  doi:{\changeurlcolor{black}\href{https://doi.org/10.1088/1361-6382/abb14e}{\detokenize{10.1088/1361-6382/abb14e}}}.

\bibitem[{Zimmermann} and {Szedenits}(1979)]{ZimmSzed1979:GrvWRtPrRBSMAppPl}
{Zimmermann}, M.; {Szedenits}, Jr., E.
\newblock {Gravitational waves from rotating and precessing rigid bodies -
  Simple models and applications to pulsars}.
\newblock {\em Physical Review D} {\bf 1979}, {\em 20},~351.
\newblock
  doi:{\changeurlcolor{black}\href{https://doi.org/10.1103/PhysRevD.20.351}{\detokenize{10.1103/PhysRevD.20.351}}}.

\bibitem[{Owen} \em{et~al.}(1998){Owen}, {Lindblom}, {Cutler}, {Schutz},
  {Vecchio}, and {Andersson}]{OwenEtAl1998:GrvWvHYRpRttNtS}
{Owen}, B.J.; {Lindblom}, L.; {Cutler}, C.; {Schutz}, B.F.; {Vecchio}, A.;
  {Andersson}, N.
\newblock {Gravitational waves from hot young rapidly rotating neutron stars}.
\newblock {\em Physical Review D} {\bf 1998}, {\em 58},~084020,
  \href{http://xxx.lanl.gov/abs/gr-qc/9804044}{{\normalfont [gr-qc/9804044]}}.
\newblock
  doi:{\changeurlcolor{black}\href{https://doi.org/10.1103/PhysRevD.58.084020}{\detokenize{10.1103/PhysRevD.58.084020}}}.

\bibitem[{Van Den Broeck}(2005)]{VanD2005:GrvWSpNnxFPrNtS}
{Van Den Broeck}, C.
\newblock {The gravitational wave spectrum of non-axisymmetric, freely
  precessing neutron stars}.
\newblock {\em Classical and Quantum Gravity} {\bf 2005}, {\em 22},~1825,
  \href{http://xxx.lanl.gov/abs/gr-qc/0411030}{{\normalfont [gr-qc/0411030]}}.
\newblock
  doi:{\changeurlcolor{black}\href{https://doi.org/10.1088/0264-9381/22/9/022}{\detokenize{10.1088/0264-9381/22/9/022}}}.

\bibitem[{Pitkin} \em{et~al.}(2017){Pitkin}, {Isi}, {Veitch}, and
  {Woan}]{PitkEtAl2017:NSmCTrSrCntGrvWP}
{Pitkin}, M.; {Isi}, M.; {Veitch}, J.; {Woan}, G.
\newblock {A nested sampling code for targeted searches for continuous
  gravitational waves from pulsars}.
\newblock {\em arXiv} {\bf 2017},
  \href{http://xxx.lanl.gov/abs/1705.08978}{{\normalfont
  [arXiv:gr-qc/1705.08978]}}.

\bibitem[{Cutler} and {Schutz}(2005)]{CutlSchu2005:GnrFsMlDtMGrvWP}
{Cutler}, C.; {Schutz}, B.F.
\newblock {Generalized $\mathcal{F}$-statistic: Multiple detectors and multiple
  gravitational wave pulsars}.
\newblock {\em Physical Review D} {\bf 2005}, {\em 72},~063006,
  \href{http://xxx.lanl.gov/abs/gr-qc/0504011}{{\normalfont [gr-qc/0504011]}}.
\newblock
  doi:{\changeurlcolor{black}\href{https://doi.org/10.1103/PhysRevD.72.063006}{\detokenize{10.1103/PhysRevD.72.063006}}}.

\bibitem[{Pitkin} \em{et~al.}(2018){Pitkin}, {Messenger}, and
  {Fan}]{PitkEtAl2018:HrrByMDtCnGrvWEP}
{Pitkin}, M.; {Messenger}, C.; {Fan}, X.
\newblock {Hierarchical Bayesian method for detecting continuous gravitational
  waves from an ensemble of pulsars}.
\newblock {\em Physical Review D} {\bf 2018}, {\em 98},~063001,
  \href{http://xxx.lanl.gov/abs/1807.06726}{{\normalfont
  [arXiv:astro-ph.IM/1807.06726]}}.
\newblock
  doi:{\changeurlcolor{black}\href{https://doi.org/10.1103/PhysRevD.98.063001}{\detokenize{10.1103/PhysRevD.98.063001}}}.

\bibitem[{Wolfram Research{,} Inc}()]{Mathematica}
{Wolfram Research{,} Inc}.
\newblock Mathematica, {V}ersion 12.0.
\newblock Champaign, IL, 2019.

\end{thebibliography}
\end{document}